# A BCS-GDE Algorithm for Multi-objective Optimization of Combined Cooling, Heating and Power Model

Jiaze Sun
School of Computer Science and Technology
Xi'an University of Posts and Telecommunications
Xi'an, China
sunjiaze@xupt.edu.cn

Jiahui Deng
School of Information Science and Technology
Northwest University
Xi'an, China
dengjiahui@stumail.nwu.edu.cn

Yang Li
School of Electrical Engineering
Northeast Electric Power University
Jilin, China
liyang@neepu.edu.cn

Shuaiyin Ma
Research Institute of Industrial Internet
Xi'an University of Posts and Telecommunications
Xi'an, China
Shuaiyinma@mail.nwpu.edu.cn

Nan Han
School of Computer Science and Technology
Xi'an University of Posts and Telecommunications
Xi'an, China
1826995691@stu.xupt.edu.cn

*Abstract*—District energy systems can not only reduce energy consumption but also set energy supply dispatching schemes according to demand. In this paper, the combined cooling heating and power economic emission dispatch (CCHPEED) model is established with the objective of economic cost, primary energy consumption, and pollutant emissions, as well as three decision-making strategies, are proposed to meet the demand for energy supply. Besides, a generalized differential evolution with the best compromise solution processing mechanism (BCS-GDE) is proposed to solve the model, also, the best compromise solution processing mechanism is put forward in the algorithm. In the simulation, the resource dispatching is performed according to the different energy demands of hotels, offices, and residential buildings on the whole day. The simulation results show that the model established in this paper can reduce the economic cost, energy consumption, and pollutant emission, in which the maximum reduction rate of economic cost is 72%, the maximum reduction rate of primary energy consumption is 73%, and the maximum reduction rate of pollutant emission is 88%. Concurrently, BCS-GDE also has better convergence and diversity than the classic algorithms.

## I. INTRODUCTION

Combined cooling heating and power (CCHP) system is used to provide distributed energy usually, it provides the energy required by the buildings on demand. CCHP system can reduce the energy loss in transmission, and improve the energy supply efficiency[1]. The conversion rate of fuel to available energy is more than 80%. In the past related work, most of the CCHP problems were modeled for the single-objective of economic[2], but since there are many influencing factors and issues that need to be considered, combined cooling heating and power economic emission dispatch (CCHPEED) has been paid more attention. In the current study, a series of multi-objective optimization models have been proposed according to different regions and energy demands. In view of this problem, how to establish a reasonable decision-making strategy model according to the actual situation, and how to find the optimal resource dispatching scheme to meet the conditions have become the current research focus.

Recently, the multi-objective CCHP system research has also been widely concerned by scholars. Tezer et al.[3] solved the problems of minimizing costs and carbon dioxide emissions in hybrid systems. Aiming at investment in distributed heating and power supply systems, a multi-objective optimization model was proposed[4], which used to minimize the total economic cost and emissions, the model was applied to the city to analyze the multi-objective problems and solutions. Dorotić et al.[5] proposed an hour-based multi-objective optimization district heating and cooling model, which can define the supply capacity including heat storage capacity and its annual operation. An optimization model with three objectives including economic cost, primary energy consumption, and pollutant emissions was established and applied in five cities, respectively[6], and the optimal operation scheme algorithm was used to determine the optimal operation mode of the model. Hu et al.[7] proposed a stochastic multi-objective optimization model, the probability constraints were added to the stochastic model to optimize CCHP operation strategies under different climatic conditions. Dorotić et al.[8] carried out a multi-objective optimization of the district heating system. The model can optimize the hourly running time of the annual time range and optimize the scale of supply capacity including storage. Although numerous models have been proposed to optimize resource scheduling, few works analyzed the demands and price of primary energy in different periods of a whole day, this is not very practical. The existing literature rarely provides different system models for different conditions.

At present, there are many methods used to solve multi-objective problems, For some complex multi-objective optimization problems, the heuristic algorithm can solve them more effectively and PSO and GA are applied to optimize the models by[9][10][11].

In the above literature, most of them used the weight function to convert multi-objective into the single-objective models when performing multi-objective calculations. The results of the simulation do not reflect the unique nature of multi-objective problems[5][8][12][13]. While the existence of a non-dominated solution can provide different resource scheduling options. Most of the algorithms have no specific indicator to measure the quality of the optimization and the performance of the algorithm.

## II. Formulation of CCHP model

In this section, a CCHP model is established. The CCHP demand of the system are provided by the power grid, power generation units (PGU), and boilers. The fuel is supplied to the generator, which produces power and waste heat. The power can be used by the building directly, also, the power demand can be met by purchasing from the grid. The recovered waste heat is used for refrigeration or heating to meet the cooling and heating demands of the buildings, and the boilers can be used to support the demand too.

Fig.1 is a network model based on the energy flow of the CCHP system. The nodes 9, 10, and 11($E_d$, $Q_{c\_d}$, $Q_{h\_d}$) in the figure are the power, cooling, and heat demands of the system. Node 1 is a concept node, which represents the total energy required by the system.

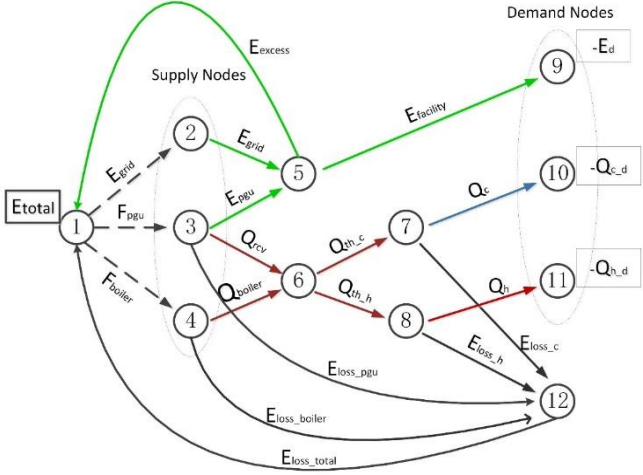

Node 1: Total energy required   Node 2: Power Grid
Node 3: PGU   Node 4: Boiler
Node 5: Electric energy provided by PGU and power grid
Node 6: Thermal energy provided by PGU and boiler
Node 7: CHP cooling components   Node 8: CHP heating components
Node 9: Electric energy demand   Node 10: Cooling energy demand
Node 11: Heating energy demand   Node 12: Total energy loss

Fig 1 Network flow model of a typical CCHP model

The model with all three types of equipment is as follows section. The second decision-making strategy is when PGU is shut down, and the third decision-making strategy is when the boiler is shut down.

### A. Objective functions

In the CCHP model built in this section, we consider three objectives. The first objective ($f_{cost}$) is to minimize the total cost of the system in T periods; the second objective ($f_{PEC}$) is to minimize the primary energy consumption in T periods; the third objective ($f_{CDE}$) is to minimize the carbon dioxide emissions of the system in T periods. The objective functions are as follow:

$$Min \ f_{cost} = \sum_{t=1}^{T}\sum_{i}(C_{fuel,i,t})$$
$$= \sum_{t=1}^{T}\{C_{el}E_{grid}(t) + C_{f\_pgu}E_{pgu}(t) + C_{f\_boiler}Q_{boiler}(t)/\eta_{boiler}\} \quad (1)$$

$$Min \ f_{PEC} = \sum_{t=1}^{T}\sum_{i}(F_{fuel,i,t})$$

$$= \sum_{t=1}^{T}\{ECF_{PEC}E_{grid}(t) + FCF_{PEC_{pgu}}(a*E_{pgu}(t)+b)\}$$
$$+ \sum_{t=1}^{T}\{FCF_{PEC_{boiler}}Q_{boiler}(t)/\eta_{boiler}\} \quad (2)$$

$$Min \ f_{CDE} = \sum_{t=1}^{T}\sum_{i}(E_{co2,i,t})$$
$$= \sum_{t=1}^{T}\{ECF_{CDE}E_{grid}(t) + FCF_{CDE\_pgu}E_{pgu}(t)\}$$
$$+ \sum_{t=1}^{T}\{FCF_{CDE\_boiler}Q_{boiler}(t)/\eta_{boiler}\} \quad (3)$$

where $t \in \{1,2,...,T\}$; in the objective of minimizing costs, the cost mainly consists of purchasing power from the power grid, PGU energy consumption cost, and boiler energy consumption cost. $C_{el}$, $C_{f_{pgu}}$ and $C_{f_{boiler}}$ respectively represent the cost of purchasing 1kwh electricity, the fuel cost of generating 1kwh energy in PGU, and the fuel cost of generating 1kwh energy in the boiler; $\eta_{boiler}$ represents the energy conversion efficiency of the boiler. In the objective of minimizing primary energy consumption, the energy consumption mainly consists of the power purchased by the grid, the energy consumed by PGU, and the energy consumed by the boiler. $ECF_{PEC}$, $FCF_{PEC\_pgu}$ and $FCF_{PEC\_boiler}$ refer to the conversion factor of primary energy for purchasing power, the primary energy conversion factor of fuel used in PGU, and the primary energy conversion factor of fuel used in the boiler; $a$ and $b$ are fuel electric energy conversion parameters. In the objective of minimizing carbon dioxide emissions, the emission mainly consists of: the emission of pollutants from the power grid, the emission of pollutants from PGU conversion fuel, and the emission of pollutants from boiler combustion. $ECF_{CDE}$, $FCF_{CDE\_pgu}$ and $FCF_{CDE\_boiler}$ represent the carbon dioxide emission conversion factor of power, the emission conversion factor of PGU fuel, and the emission conversion factor of boiler fuel, respectively.

### B. Constraints

In this model, a series of constraints need to be met to optimize the objective functions, including energy conservation constraints, fuel conversion constraints, and energy efficiency constraints. The constraint formula is as follows. The energy conservation constraints of nodes 3, 4, 5, 6, 7, 8, and 12 in Fig.1 are as follows:

$$E_{pgu}(t) + Q_{rcv}(t) + Energy_{loss\_pgu}(t) - F_{pgu}(t) = 0 \quad (4)$$

$$Q_{boiler}(t) + Energy_{loss\_boiler}(t) - F_{boiler}(t) = 0 \quad (5)$$

$$E_{excess}(t) + E_{facility}(t) - E_{grid}(t) - E_{pgu}(t) = 0 \quad (6)$$

$$Q_{th\_cool}(t) + Q_{th\_heat}(t) - Q_{rcv}(t) - Q_{boiler}(t) = 0 \quad (7)$$

$$Q_{cool}(t) + Energy_{loss\_c}(t) - Q_{th\_cool}(t) = 0 \quad (8)$$

$$Q_{heat}(t) + Energy_{loss\_h}(t) - Q_{th\_heat}(t) = 0 \quad (9)$$

$$Energy_{loss\_total}(t) - Energy_{loss\_pgu}(t) - Energy_{loss\_boiler}(t) - Energy_{loss\_c}(t) - Energy_{loss\_h}(t) = 0 \quad (10)$$

Equations (4) and (5) are the conservation conditions of converting the energy generated by PGU and boiler to available energy, respectively. Equations (6) and (7) respectively represent the conservation relationship between the total power generated, the total heat generated, and the energy output. Equations (8) and (9) respectively represent the conservation relationship between the energy generated by the cooling component, the heating component, and the output energy. Equation (10) shows the conservation relation of total energy loss of the system.

## III. OPTIMIZATION METHOD

### A. BCS-GDE

Since the model proposed in this paper is a multi-objective optimization problem with constraints, the generalized differential evolution with the best compromise solution processing mechanism (BCS-GDE) algorithm will be proposed to solve it.

BCS-GDE is an extension of differential evolution[14] (DE) and the third evolution step of generalized differential evolution[15] (GDE3). DE can only be used to solve single-objective optimization problems, while BCS-GDE can be used to solve multi-objective optimization problems with constraints. BCS-GDE is different from GDE3 in that BCS-GDE can find the best compromise in the optimal solution set. The calculation flow chart of BCS-GDE is shown in Fig.2:

Fig 2 Flow chart of BCS-GDE

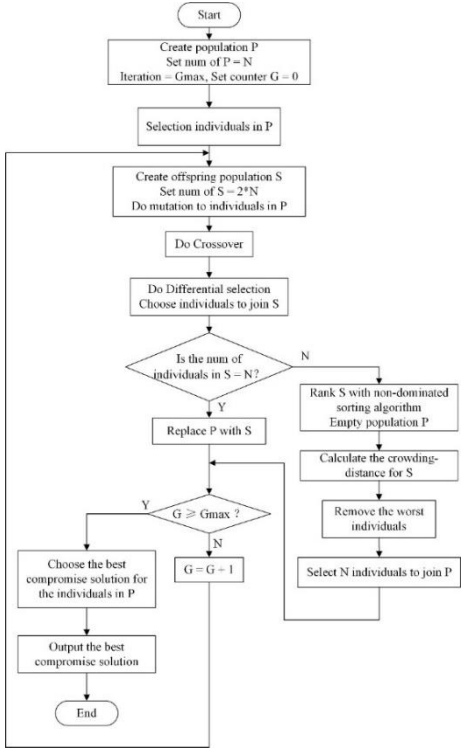

### B. The choice strategy of the best compromise solution

Since the model proposed in this paper is a multi-objective optimization problem with constraints, the generalized differential evolution with the best compromise solution processing mechanism (BCS-GDE) algorithm will be proposed to solve it.

The solution to the multi-objective CCHP model is a set of optimal solutions. These solutions dominate each other and cannot determine which is the optimal solution. Therefore, there is no systematic and standard way to choose the scheduling scheme. In order to solve this problem, the author proposes an optimal compromise solution selection strategy, which can select the approximate optimal solution from a set of optimal solutions.

Generally speaking, the most ideal situation is that the three objective functions in the model reach the minimum values simultaneously, and the ideal point in the solution space is (0, 0, 0), but it is not likely to occur in practical problems. Therefore, in the process of choosing the best compromise solution, the Euclidean distances between the non-dominated solutions and the ideal point are calculated, the solution with minimum distance is chosen as the approximate optimal solution. The calculation method is as follows (10):

$$Dist(X,Y) = \sqrt{\sum_{i=1}^{n}(x_i - y_i)^2} \quad (10)$$

where $X$ and $Y$ represent two n-dimensional vectors respectively, $X(x_1, x_2, \dots, x_n)$, $Y(y_1, y_2, \dots, y_n)$. In this paper, $Y(y_1, y_2, \dots, y_n) = (0, 0, \dots, 0)$. The specific solution flow of approximate optimal solution is shown in Fig.3:

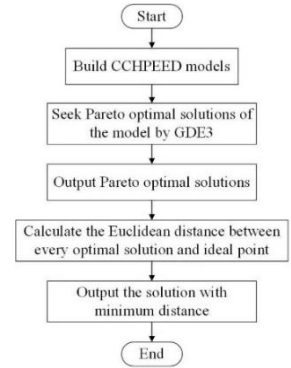

Fig 3 Flow chart of best compromise solution selection

## IV. SIMULATION AND DISCUSSION

### A. Data analysis

In the simulation, the energy supply is provided for three types of buildings include hotels, offices, and residential buildings, the energy demands are different. The input data and parameter information are summarized, and the different application scenarios are analyzed in details.

*1) Input Data*

It is assumed that the virtual buildings used for simulation are all located in Xi'an, Shaanxi, China. There are four independent buildings in the hotel area with a total area of $202,768 m^2$, four independent buildings in the office area with a total area of $197,568 m^2$, and eight independent buildings in the residential area with a total area of $199,064 m^2$ [16]. In the CCHP system, the electricity price on the power grid and the natural gas are shown in Table 1. The electricity prices of every type of buildings are different in each power consumption period, as well as the price of natural gas, all the data are collected locally. In the case calculation, natural gas is used for PGU and boiler, the power grid, and PGU supply power. Table 2 is the factor of converting primary energy to available energy, and every type of energy has different conversion factors. The CCHP system produces emissions when it works. Table 3 shows the carbon dioxide conversion factors of electric energy and natural gas. Table 4 shows the constraints of conversion and efficiency of CCHP in the process of energy output and transmission.

In the simulation, the reference system means that the energy supply without the CCHP system. The power of the reference system is purchased from the power grid, while the heating and cooling demand are supplied by using natural gas as the fuel through heating components and cooling components.

Table 1 Electricity and natural gas prices (Yuan/kWh).

|  | Electricity price | | Natural Gas price |
|---|---|---|---|
|  | Commercial buildings | Residential buildings |  |
| Average time | 0.87 | 0.5 | 0.22 |
| Peak load time | 1.305 | 0.65 | 0.22 |
| Low load time | 0.435 | 0.45 | 0.22 |

Table 2 Site-to-primary energy conversion factor[6].

| Fuel type | Conversion factor |
|---|---|
| Electricity | 3.336 |
| Natural Gas | 1.047 |

Table 3 Emission conversion factors for electricity and natural gas.

| Emission | Electricity(g/kWh) | Natural Gas(g/kWh) |
|---|---|---|
| $CO_2$ | 203.74 | 200 |

Table 4 Conversion and efficiency constraints[6].

|  | Symbol | Value |
|---|---|---|
| Fuel-to-electric-energy conversion parameter | $a$ | 2.67 |
| Fuel-to-electric-energy conversion parameter | $b$ | 11.43 |
| Fuel-to-thermal-energy conversion efficiency of PGU | $\eta_{pgu\_th}$ | 0.51 |
| Boiler efficiency | $\eta_{pgu\_th}$ | 0.9 |
| Total efficiency of the cooling components | $\eta_{pgu\_th}$ | 0.7 |
| Total efficiency of the heating components | $\eta_{pgu\_th}$ | 0.85 |

*2) Scenario analysis*

In the simulation, three kinds of buildings are designed to verify the effectiveness of the CCHP problem by using three actual scenes. As shown in Fig.4, simulations carried out under energy requirements in summer, winter, and transition season. In winter, less cooling is needed, and the demand for heating is the least in summer. While the demand has different rules according to buildings, the demand of the hotel is relatively large and stable, offices have a larger heating and cooling demand in normal working hours, residential buildings produce a certain demand according to residents' living habits. In transition seasons, there are both heating demand and cooling demand. Table 5 shows the peak energy demand of various types of buildings.

Table 5 The electricity, cooling, and heating load of the buildings (kW).

|  | Hotel | Office | Residential buildings |
|---|---|---|---|
| Electricity load | 3070 | 3198 | 4166 |
| Cooling load | 5400 | 7056 | 6145 |
| Heating load | 7657 | 7050 | 7080 |

*3) Simulation settings*

The optimization models of the CCHP system have been established in section II, in this section, the process of the simulation will be addressed. Three decision variables are list in Table 6, the objective function can be optimized by scheduling the values of these three variables.

Table 6 Decision variables of the models.

| Decision variables | Description |
|---|---|
| X1 | Electricity purchased from the grid |
| X2 | Natural gas consumption by PGU |
| X3 | Natural gas consumption by the boiler |

To unify the simulation environment, all parameter settings are the same during the simulation, the parameters of BCS-GDE are list in Table 7.

Table 7 Parameters of BCS-GDE.

| Parameter | Value |
|---|---|
| Population size | 100 |
| Maximum iterations | 250 |
| CR | 0.5 |
| F | 0.5 |

*B. Results and discussion*

The CCHP system models and BCS-GDE algorithm proposed in this paper are all implemented in jMetal 4.5 with JDK 1.8, and the computer environment is 2.8 GHz Intel Core i7, 8 GB RAM. Various algorithms for comparative experiments have also been implemented in jMetal and run under the same environment and conditions. When solving multi-objective problems, the average execution time of getting the Pareto approximate solution set of the BCS-GDE algorithm is 312ms, which is shorter than other evolutionary algorithms used for comparison in simulations.

The scenario in this section is in the transitional season, referring to spring and autumn. From Fig.4, it can be seen that in these seasons, the demand for power of all kinds of buildings is more than that of other seasons, and there is a demand for cooling and heating in this season; the demand for energy in offices fluctuates obviously in a day, while that in hotels and residential buildings is relatively flat. BCS-GDE algorithm obtains 100 non-dominated solutions according to the different energy demands of each day. In the simulation, the best compromise solution will be selected.

Fig.5 states the change rate of economic cost, PEC, and CDE between the reference system and the strategies, respectively. The improvement rate represents the percentage reduced by the proposed strategies compared with the reference system in the objective of economic cost, PEC, and CDE, respectively.

Case 1 represents the CCHP system, case 2 represents the decision-making strategy when PGU is shut down, case 3 represents the decision-making strategy when the boiler is shut down. The above definition applies to all simulations.

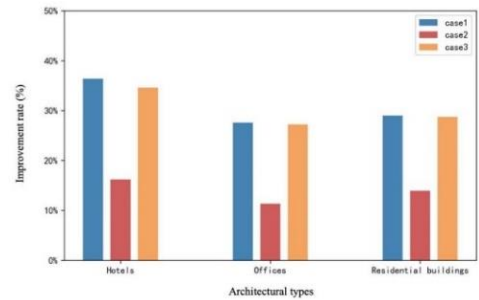

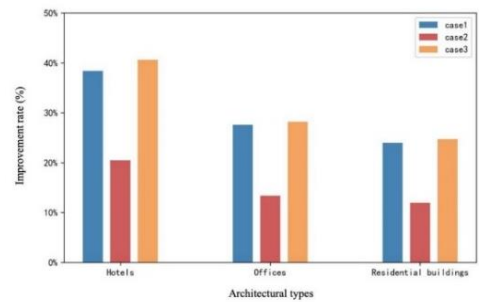

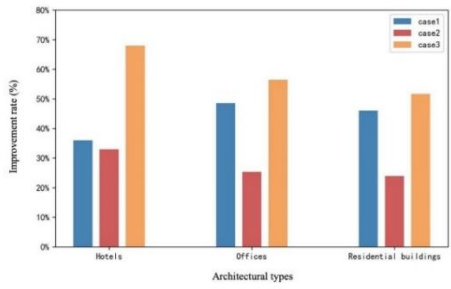

Fig 5 Change rate of cost, PEC and CDE in transitional season

From the overall point of view in Fig.5, compared with the three objective values of the reference system, the three strategies have a significant improvement, and case 2 has a smaller improvement compared with the other two strategies. As one of the core components of the CCHP system, PGU can provide both power and heating. The system with PGU has a higher energy saving rate than that without PGU.

In the objective of economic cost, the change rate of case 1 is from 28% to 36%, with the most obvious improvement, which means that compared with the reference system, case 1 saves at least 28% of the economic cost, while case 2 changes from 12% to 15%. For the objective of PEC, case 3 has the most obvious change rate, from 25% to 41%, saving more than 25% of primary energy consumption for the system. For CDE, case 3 has a change rate of 52% to 69% compared with the reference system, which can reduce more than half of the pollutant emissions of the system.

Both in the objective of PEC and CDE, case 3 performance a better improvement than case 1, in the objective of economic cost, both of them have a similar result. In this scenario, case 3 will be accepted by the most suitable scheme to supply energy.

To verify the effectiveness and superiority of the proposed algorithm BCS-GDE in solving the CCHPEED, the simulation takes the peak energy demand of residential buildings in Table 5 as the rated energy demand, different algorithms are chosen to solve the CCHPEED model, and the Pareto approximate fronts are compared in Fig.6.

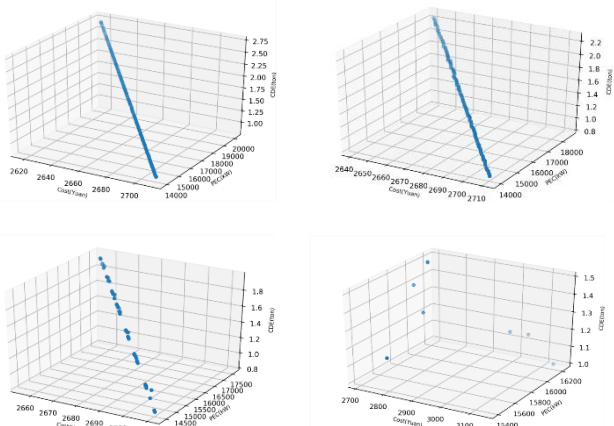

Fig 6 Pareto approximation front of BCS-GDE, OMOPSO, NSGA-II and SPEA2(from the upper left to the lower right)

It can be seen from Fig.6 that in the same experimental environment, BCS-GDE can find the non-dominated solution with wide and evenly distributed, and the non-dominated solution found by OMOPSO has certain universality but poor convergence compared with BCS-GDE. The non-dominated solution obtained by NSGA-II has uneven distribution and few solutions, while SPEA2 can only find very few non-dominated solutions with poor convergence.

Hypervolume (HV) and spread ($\triangle$) indicators are selected to evaluate the comprehensive performance and diversity of BCS-GDE and other algorithms in the simulation. To ensure the objective evaluation, each algorithm runs 20 times to obtain 20 indicator values respectively, and the maximum, minimum, and average values of each indicator value are shown in Table 8.

The larger the HV indicator value is, the better the comprehensive performance of the algorithm is, the smaller the spread indicator value is, the more extensive and uniform the solution can be distributed on the Pareto front. According to Table 8, the HV value of BCS-GDE is 0.33, which is larger than that of the other algorithms. The spread indicator value of BCS-GDE is 0.15, which is the smallest compared with other algorithms. It shows that BCS-GDE not only has a more extensive and uniform distribution of non-dominated solutions, but also has better comprehensive performance compared with OMOPSO, NSGA-II, and SPEA2.

To verify the statistical results of the algorithms, the Wilcoxon signed-rank test is used for the performance difference of pair-wise comparison algorithms. The following hypothesis is proposed:

$H_0$ :BCS-GDE has a significant improvement over the algorithm

Table 9 shows the statistical results, the p-value is considered to reject $H_0$ or not. And all the p-values are less than the significance level α, and the hypothesis is accepted, $H_0$: BCS-GDE has a significant improvement over OMOPSO, NSGA-II, and SPEA2.

Table 9 Wilcoxon signed-rank test results with significance level α = 0.001.

| Methods | p-value (HV) | p-value ($\triangle$) |
|---|---|---|
| **BCS-GDE vs OMOPSO** | 0.001 | 0.0001 |
| **BCS-GDE vs NSGA-II** | 0.0001 | 0.0001 |
| **BCS-GDE vs SPEA2** | 0.0001 | 0.0001 |

## V. CONCLUSION

In this paper, the multi-objective mathematical optimization model of CCHPEED is established and optimized by BCS-GDE. Three decision strategies of CCHPEED models are proposed for different energy supply demands as well. In the simulation, the system provided energy to hotels, offices, and a detailed data analysis is carried out. That is, the models established in this paper can greatly reduce the economic cost, primary energy consumption, and carbon dioxide emissions. In terms of cost, PEC, and CDE objectives, the model can save 88%, 73%, and 72% of the reference system at most. In the simulation, the performance of algorithms is compared and analyzed, BCS-GDE shows better effectiveness and superiority than other classical algorithms. Also, the simulation shows BCS-GDE has a significant improvement over the classical algorithms. But considering the development of clean energy, in future work, the adjunction of solar energy, wind energy, natural gas, and fuel cell may perform better in energy scheduling.

Table 8 Quality evaluation of OMOPSO, NSGA-II, SPEA2, and BCS-GDE

| Indicator | OMOPSO | | | NSGA-II | | | SPEA2 | | | BCS-GDE | | |
|---|---|---|---|---|---|---|---|---|---|---|---|---|
| | max | min | ave | max | min | ave | max | min | ave | max | min | ave |
| HV | 0.33 | 0.32 | 0.32 | 0.32 | 0.30 | 0.31 | 0.33 | 0.27 | 0.31 | 0.33 | 0.33 | **0.33** |
| △ | 0.27 | 0.15 | 0.20 | 1.24 | 0.96 | 1.13 | 1.32 | 0.95 | 1.15 | 0.18 | 0.13 | **0.15** |

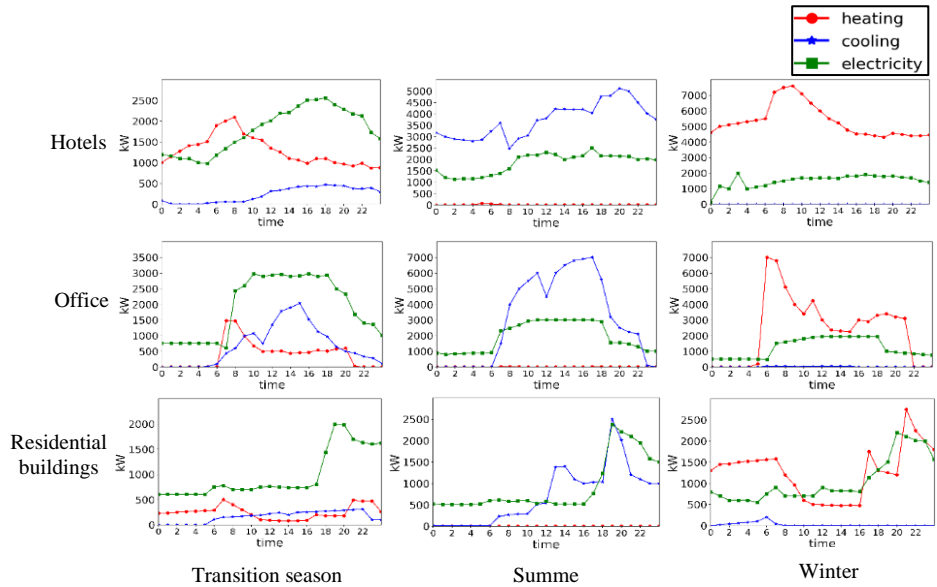

Fig 4 Daily energy load of buildings in CCHHP system

ACKNOWLEDGMENT

The work is supported by the National Natural Science Foundation of China (Grant No. 61876138), the Key R & D Project of Shaanxi Province (2020GY-010), the Industrial Research Project of Xi'an (2019218114GXRC017CG018-GXYD17.10), and the Special Fund for Key Discipline Construction of General Institutions of Higher Learning from Shaanxi Province.